\title{Scala-Gopher:  CSP-style programming techniques with idiomatic Scala. }
\author{
        Ruslan Shevchenko \\
                ruslan@shevchenko.kiev.ua\\
        Kiev, Ukraine \\
         scala-2016 OSS Talk
}
\date{\today}

\documentclass[12pt]{article}
\usepackage{listings}
\usepackage{amsmath}
\usepackage{amssymb}
\usepackage{pgf}
\usepackage{tikz}
\usepackage{fancyvrb}
\usetikzlibrary{arrows,automata}
\usepackage[backend=bibtex]{biblatex}
\newcommand{\To}{\Rightarrow}

\begin{document}
\maketitle

\section{Introduction}

 Scala-gopher is a library-level implementation of process algebra [Communication Sequential Processes, see \cite{Hoare85communicatingsequential} as ususally enriched by $\pi$-calculus \cite{Milner:1992:CMP:162037.162038} naming primitives] in scala. In addition to support of a 'limbo/go-like' \cite{Inferno:Limbo}  \cite{golang} channels/goroutine programming style scala-gopher provide set of operations following typical scala idiomatic. 

    At first, let's remind the fundamentals of a CSP model. The primary entities of this model are channels, coroutines and selectors. Coroutines are lightweight threads of execution, which can communicate with each other by passing messages between channels. Channels can be viewed as blocked multiproducer/multiconsumer queues. Sending message to unbuffered channel suspend producing coroutines until the moment when this message will have been read by some consumer. Buffered channels transfer control flow between sinks not on each message, but when internal channel buffer is full.  
In such way, communication via channel implicitly provides flow control functionality.  At last, a selector statement is a way of coordination of several communication activities: like Unix select(2) system call, select statement suspends current coroutines until one of the actions (reading/writing to one of the channels in selector) will be possible.

   Let's look at the one simple example:
\begin{Verbatim}[fontsize=\small]
 def nPrimes(n:Int):Future[List[Int]]=
 {
    val in = makeChannel[Int]()
    val out = makeChannel[Int]()
    go {
      for(i <- 1 to n*n) out.write(i)
    }
    go {
      select.fold(in){ (ch,s) =>
        s match {
          case p:ch.read => out.write(p)
                            ch.filter(_ % p != 0)
        }
      }
    }
    go {
      for(i <- 1 to n) yield out.read
    }
 }
\end{Verbatim}
  Here two channels and three goroutines are created.  The first coroutine just generates consecutive numbers and send one to channel \verb|in|, second - accept this sequence as the initial state and for each number which has been read from state channel, write one to \verb|out| and produce next step by filtering previous. The result of the fold is the \verb|in| channel whith applied filters for each prime. The third coroutine just maps range to values to receive a list of first \verb|n| primes in Future.

  If we look at the sequence of steps during code evaluation, we will see at first generation of
number, then checks in filters and then if a given number was prime - final output.  Note, that goroutine is different from JVM thread of execution: sequential code chunks are executed in configurable executor service; switching between chunks does not use blocking operations.

\section{Implementation of base constructs }

\subsection{Go: Translations of hight-order functions to asynchronous form.}

 The main entity of CSP is a 'process' which can be viewed as a block of code which handles specific events.  In Go CSP processes are represented as goroutines (aka coroutines). 

 \verb|go[X](x:X):Future[X]| is a think wrapper arround  SIP-22 async/await which do some preprocessing before async transformation:
\begin{itemize}
 \item do transformation of hight-order function in async form.

  Let $f(A \To B)\To C$ is a hight-order function, which accepts other function 
   $g:A \To B$ as parameter. 
  Let's say that $g$ in $f$ is {\i invocation-only } if $f$ not store $g$ in memory outside of $f$ scope and not return $g$ as part of return value. Only one action which $f$ can do with $g$ is invocation or passing as a parameter to other invocation-only function. If we look at Scala collection API, we will see, that near all hight-order functions there are invocation-only.

  Now, if we have $g$ which is invocation-only in $f$, and if we have function $g' : (A \To Future[B])$ let build function $f':(A\To Future[B])\to Future[C]$ that if $await(g')==await(g)$ then $await(f'(g'))==f(g))$ in next way
  \begin{itemize}
    \item $f'$ translated to $await(\makebox{transformed-body}(f))$
    \item $g(x)$ inside $f$ translated to $await(g'(x))$
    \item $h(g)$ translated to $await(h'(g'))$ if $g$ is invocation-only in $h$.
  \end{itemize}

  Scala-gopher contains asynchronious variants of predefined functions from Scala collection API, so it is possible to use asynchronious expressions inside a loop. For example, next code:

\begin{Verbatim}[fontsize=\small]
  go { 
      for(i <- 1 to n) yield out.read
  }
\end{Verbatim}

  is transformed to

\begin{Verbatim}[fontsize=\small]
 async{ await {
   (1 to n).mapAsync(i => async{ await{ out.aread } } )
 } }
\end{Verbatim}

  which after simplification step become

\begin{Verbatim}[fontsize=\small]
   (1 to n).mapAsync(i => out.aread)
\end{Verbatim}

 Using this approach allows overcoming the inconvenience of async/await by allowing programmers use hight-order functions API inside asynchronous expression. Also, it is theoretically possible to generate asynchronous variants of API methods by transforming TASTY representation of AST of synchronous versions. Thr similar technique is implemented in Nim \cite{Nim} programming language, where we can to generate both synchronious and asynchronious variants of a function from one definition.
  
 \item do transformation of defer statement. This is just an implementation of error handling mechanism.

\end{itemize}

\subsection{Channels: callbacks organized as waits}

  Channels in CSP are two-sided pipes between processes; Channels messages not only pass information between goroutines but also coordinate process execution.
  In scala-gopher appropiative entities (\verb|Input[A]| for reading and \verb|Output[A]| for writing) implemented in fully asynchniously manner with help of callback-based interfaces:

\begin{Verbatim}[fontsize=\small]
trait Input[A]
{

   def  cbread[B](f:
            ContRead[A,B]=>Option[
                    ContRead.In[A]=>Future[Continuated[B]]
            ],
            ft: FlowTermination[B]): Unit
   ....
}
\end{Verbatim}

  Here we can read argument type as protocol where each arrow is a step: 
    $f$ is called on opportunity to read and
      $ContRead[A,B] \To Option[ContRead.In[A] \To Future[B]]$ means that when reading is 
   possible, we can ignore this opportunity (i.e. return None) or return handler which will
   consume value (or \verb|end-of-input| or few other special cases) and return future to the next 
   computation state.

  Traditional synchronious API  (i.e. method $read:\To A$) can be used inside \verb|go| and \verb|async| statements; from 'normal' code we can use asynchronious variant: $aread: \To Future[A]$.

  Output interface is similar:
\begin{Verbatim}[fontsize=\small]
trait Output[A] 
{

  def  cbwrite[B](f: ContWrite[A,B] => Option[
                   (A,Future[Continuated[B]])
                  ],
                  ft: FlowTermination[B]): Unit
  
}
\end{Verbatim}
  Here $f$ is called on opportunity to write and when we decide to use this opportunity, we
 must provide actual value to write and next step in the same way as with \verb|Input|.

  Inputs and outputs are composable as can be expected in a functional language and equipped by 
the usual set of stream combinators: filter, map, zip, fold, etc.
 
  Channel is a combination of input and output. In addition to well-known buffered and unbuffered 
kinds of channels, scala-gopher provide some extra set of channels with different behavior and performance characteristics,
 such as the channel with growing buffer (a-la actor mailbox) for connecting loosely coupled processes or one-time channel based on \verb|Promise|, which is automatically closed after sending one message.

\subsection{Selectors: process composition as event generation}

 Mutually exclusive process composition (i.e. deterministic choice: $(a \to P)\square(b\to Q)$ in original Hoar notation )  usually represented in CSP=based languages as \verb|select| statement, which look's like ordinary switch. In a typical \verb|Go| program exists often repeated code pattern: select inside endless loop inside go statement.

\begin{Verbatim}[fontsize=\small]
go {
  for{
    select{
     case c1 -> x :
           ......... // P
     case c2 <- y :
           ........  // Q
    }
  }
}
\end{Verbatim}

  Appropriative expression in CSP syntax: $*[(c_{1} ? x \to P)\square(c_{2} ! y \to Q)]$

 Scala-gopher provide \verb|select| pseudo-object which provide set of high-order pseudo-functions over
channels, which accept syntax of partial function over channel events:

\begin{Verbatim}[fontsize=\small]
go {   
  select.forever {
    case x: c1.read => ....  //P
    case y: c2.write => ....  //Q
  }
}
\end{Verbatim}
   
  or version which must not be wrapped by \verb|go| stamenet:

\begin{Verbatim}[fontsize=\small]
  select.aforever {
    case x: c1.read => ....  //P
    case y: c2.write => ....  //Q
  }
\end{Verbatim}
   
  Under the hood, each such pseudo-function is built around a flow (sequence of \verb|Continuated[_]| which represents the step of computations optionally bound to channel event).  In unsugared form selector 

\begin{Verbatim}[fontsize=\small]
  val selector = SelectorForever()
  selector.onRead(ch)((x,ft,ec) => ... ) // P after go-transform
  selector.onWrite(ch,y)((y,ft,ec) => ... ) // Q after go-transform
  selector.run()
\end{Verbatim}

 We can maintain state inside a flow in a clear functional manner  using 'fold' family of select functions:

\begin{Verbatim}[fontsize=\small]
  def fibonacci(c: Output[Long], quit: Input[Boolean]): Future[(Long,Long)] =
     select.afold((0L,1L)) { case ((x,y),s) =>
      s match {
        case x: c.write => (y, x+y)
        case q: quit.read =>
                   select.exit((x,y))
      }
     }
\end{Verbatim}

  Here we see the special syntax for tuple state. Also note, that \verb|afold| macro assume that \verb|s match| must be the first statement in the argument pseudo-function. \verb|select.exit| is used for returning result from the flow.
 
  Events which we can check in select match statement are reading and writing of channels and select timeouts. In future we will think about extending the set of notifications - i.e. adding channel closing and overflow notifications, which are rare needed in some scenarios.

\subsection{Transputer: an entity which encapsulates processing node. }

 The idea is to have an actor-like object, which encapsulates processing node: i.e., read input data 
from the set of input ports; write a result to the set of output ports and maintaine a local 
mutable state inside.

Example:

\begin{Verbatim}[fontsize=\small]
class Zipper[T] extends SelectTransputer
{
 
   val inX: InPort[T]
   val inY: InPort[T]

   val out: OutPort[(T,T)] 

   loop {
     case x: inX.read => 
             val y = inY.read
             out write (x,y)
     case y: inY.read =>
             val x = inX.read
             out.write((x,y)) 
   }


}
\end{Verbatim}

  Having set of such objects, we can build complex systems as combination of simple ones:
  \begin{itemize}
    \item $a+b$ - parallel execution; 
    \item $replicate[A](n)$ - transputer replication, where we start in parallel $n$ instances
  of $A$. Policy for prot replication can be configured - from sharing appropriative channel by each port to distributing   or duplication of signals to ditinguish each instance.
\begin{Verbatim}[fontsize=\small]
 val r = gopherApi.replicate[SMTTransputer](10)
    ( r.dataInput.distribute( (_.hashCode % 10 ) ).
       .controlInput.duplicate().
        out.share()
    )
\end{Verbatim}
    - here in \verb|r| we have ten instances of \verb|SMTTransputer|. If we send a message to \verb|dataInput| it will be directed to one of the instances (in dependency from the value of message \verb|hashcode|). If we send a message to the \verb|controlInput|, it will be delivered to each instance; output channel will be shared between all instances.
  \end{itemize}

  Transputers can participate in error handling scenarios in the same way as actors: for each transputer, we can define recovery policy and supervisor.

\subsection{ Programming Techniques based on dynamic channels  }

 Let's outline some programming techniques, well known in Go world but not easily expressible in current mainstream Scala streaming libraries. 

\begin{itemize}
 \item Channel expression as an element of runtime state. Following this pattern allows a developer to maintain dynamics potentially recursive dataflows. 
 
 Example: 
  Imagine situation, where we need to distribute some tasks across a set of relative slow consumers and we need to spawn additional consumers on peak usage and free resources during the calm. 

\begin{Verbatim}[fontsize=\small]
 select.fold(output){ (out, s) => s match {
   case x:input.read =>
     select.once {
       case x:out.write =>
       case select.timeout =>
             control.distributeBandwidth match {
                case Some(newOut) => newOut.write(x)
                                     (out | newOut)
                case None => control.report("Can't increase bandwidth")
                                     out
             }
     }
   case select.timeout =>
     out match {
       case OrOutput(frs,snd) => snd.close
                                 frs
       case _                 => out
     }
 } }
\end{Verbatim}

 Here we can request additional channels from control and construct merged channel in the state of the fold. On read timeout, we can deconstruct merged channel back and free unused resources.

\item{ Channel-based API where client supply channel where to pass reply }

 Let we want provide API which must on request return some value to the caller. Instead of providing a method which will return a result on the stack we can provide endpoint channel, which will accept method arguments and channel where to return a result. 

 Next example illustrate this idea:

\begin{Verbatim}[fontsize=\small]
trait Broadcast[T]
{
   val listener: Output[Channel[T]]
   val messages: Input[T] 
}

class BroadcastImpl[T] extends Broadcast[T]
{

   val listener: Channel[Channel[T]] = makeChannel[Channel[T]]
   val messages: Channel[T] = makeChannel[]

   // private part
   case class Message(next:Channel[Message],value:T)

   select.afold(makeChannel[Message]) { (bus, s) =>
      s match {
         case v: message.read =>
                   val newBus = makeChannel[Message]
                   current.write(Message(newBus,v))
                   newBus
         case ch: listener.read =>          
                   select.afold(bus) { (current,s) =>
                     s match {
                       case msg:current.read =>
                               ch.awrite(msg.value) 
                               msg.next
                     }
                   } 
                   current
     } 
  }

}
\end{Verbatim}

 Broadcast provide API for creation of listeners and sending messages to all listeners.

To register listener channel for receiving notification client sends this channel to newListener
  
 The internal state contains message bus represented by a channel which is replaced during each new input message. Each listener spawns the process which reads messages from the current message bus.

\end{itemize}

\section{ Connection with other models }

 Exists many stream libraries for Scala with different sets of tradeoffs. At one side of
spectrum, we have clear streaming models like akka-streams\cite{akka-streams} with comp set
of composable operations and clear high-level functionality but lack of flexibility,
from another side - very flexible but low-level models like actors.

 Scala-gopher provides uniform API which allows build systems from different parts of spectrum:
it is possible to build dataflow graph in a declarative manner and connect one with a dynamic part.

 The reactive isolates model\cite{Prokopec:2015:ICE:2814228.2814245} is close to scala-gopher model with dynamically-grown channel buffer (except that reactive isolates support distributed case).
 Isolate here corresponds to Transputer, Channel to Output and Events to gopher Input. Channels in reactive isolates are more limited: only one isolate which owns the channel can write to it when in the scala-gopher concept of channel ownity is absent. 

 Communicating Scala Objects\cite{CSO} is a direct implementation of CSP model in Scala which allows
building expressions in internal Scala DSL, closed to origin Hoar notation with some extensions, like extended rendezvous for mapping input streams.  Processes in CSO are not lightweight: each process requires Java thread which limits the scalability of this library until lightweight threading will be implemented on JVM level.

 Subscript\cite{vanDelft:2013:DCL:2489837.2489849} is a Scala extension which adds to language new constructions for building process algebra expressions. Although extending language can afford fine-grained interconnection of process-algebra and imperative language notation in far perspective, now it makes CPA constructs a second-class citizen because we have no direct representation of process and event types in Scala type system.

\section{ Conclusion and future directions }
  
  Scala-gopher is a relatively new library which yet not reach 1.0 state, but we have the early experience
 reports from using the library for building some helper components in an industrial software project. 

 In general, feedback is positive: developers enjoy relative simple mental model and ability freely use asynchronous operations inside hight-order functions.  So, we can recommend to made conversion of invocation-only functions into async form to be available into the async library itself. 

 The area which needs more work: error handling in \verb|go| statements:  now \verb|go| return 
\verb|Future| which can hold a result of the evaluation or an exception. If we ignore statement result than we miss handling of exception; from another side, it unlikely handle errors there, because we must allow a developer to implement own error processing. The workaround is to use different method name for calling statement in the context with ignored return value, but it is easy to mix-up this two names.
 We think, that right solution can be built on language level: we can bind handling of ignored value to type by providing appropriative implicit conversion or use special syntax for functions which needs 
 
 Also, we plan to extend model by adding notifications about channel-close and channel-overflow on write side which needed in some relatively rare usage scenarious.

 Support of distributed communications can be the next logical step in future development. In our opinion, practical approach will differ from implementing location-transparency support for existing API. Rather we will think to enrich model with new channel types and mechanisms for work in explicitly distributed environment.

  Finally, we can say that CSP model can be nicely integrated with existing Scala concurrency ecosystem and have a place in existing zoo of concurrency models.  Using scala-gopher allows developers to use well establishment elegant techniques, such as dynamic recursive data flows and channel-based API-s. 

\section{Literature}

\printbibliography

\end{document}